\documentclass[a4paper,11pt]{article}
\pdfoutput=1 

\usepackage{jcappub} 

\usepackage[LGR,T1]{fontenc}
\usepackage[utf8]{inputenc}
\usepackage{textalpha}

\usepackage{subfigure}
\usepackage{enumitem}
\usepackage{caption}

\title{Decaying Dark Matter: Simulations and Weak-Lensing Forecast}

\author[]{Jonathan Hubert,}
\author[]{Aurel Schneider,}
\author[]{Doug Potter,}
\author[]{Joachim Stadel,}
\author[]{and Sambit K. Giri}

\affiliation[]{Institute for Computational Science, University of Zurich,\\Winterthurerstrasse 190, 8057 Zurich, Switzerland}

\emailAdd{aurel.schneider@uzh.ch}

\abstract{Despite evidence for the existence of dark matter (DM) from very high and low redshifts, a moderate amount of DM particle decay remains a valid possibility. This includes both models with very long-lived yet unstable particles or mixed scenarios where only a small fraction of dark matter is allowed to decay. In this paper, we investigate how DM particles decaying into radiation affect non-linear structure formation. We look at the power spectrum and its redshift evolution, varying both the decay lifetime ($\tau$) and the fraction of decaying-to-total dark matter ($f$), and we propose a fitting function that reaches sub-percent precision below $k\sim10$ h/Mpc. Based on this fit, we perform a forecast analysis for a Euclid-like weak lensing (WL) survey, including both massive neutrino and baryonic feedback parameters. We find that with WL observations alone, it is possible to rule out decay lifetimes smaller than $\tau=75$ Gyr (at 95 percent CL) for the case that all DM is unstable. This constraint improves to $\tau=182$ Gyr if the WL data is combined with CMB priors from the Planck satellite and to $\tau=275$ Gyr if we further assume baryonic feedback to be fully constrained by upcoming Sunyaev-Zeldovich or X-ray data. The latter shows a factor of 3.2 improvement compared to constraints from CMB data alone. Regarding the scenario of a strongly decaying sub-component of dark matter with $\tau\sim 30$ Gyr or lower, it will be possible to rule out a decaying-to-total fraction of $f>0.49$, $f>0.21$, and $f>0.13$ (at the 95 percent CL) for the same three scenarios. We conclude that the upcoming stage-IV WL surveys will allow us to significantly improve current constraints on the stability of the dark matter sector.}

\begin{document}
\maketitle
\flushbottom

\section{Introduction}
\label{sec:intro}
Current cosmological and astrophysical observations indicate that dark matter (DM) is present during both the very early and the late universe. Indeed, a substantial fraction of DM is required to explain the cosmic microwave background (CMB) radiation originating at the epoch of recombination \citep{Planck}. At the same time, weak-lensing shear measurements \citep{KidS, DES}, galaxy clustering statistics \citep{BOSSpaper}, individual galaxy rotation curves, or strong-lensing observations \citep{Holicow} point towards the need of a comparable DM abundance at low redshift. Combining all observational evidence therefore leads to the obvious conclusion that a hypothetical dark matter particle needs to be either stable or, at least, very long-lived.

Current limits on the DM particle half-life are at $\tau\sim 75$ Gyr or higher (depending on the used data combination from e.g. CMB, weak-lensing or cluster-count observations), which is more than five times the age of the universe \cite{Ichiki:2004vi,Enqvist15, Enqvist19, DESpaper, Audren2014}. Note, however, that this constraint can be substantially relaxed if only part of the DM is assumed to be unstable. In such a multi-component scenario, the half-life $\tau$ can be significantly shorter, remaining entirely unconstrained if only $\sim20$ percent of the DM budget is allowed to decay \citep{Poulin16}.

Recent weak-lensing observations from KiDS \citep{KidS}, DES \citep{DES}, and HSC \citep{Hikage2019} have found a surprisingly low clustering amplitude in 2-3 $\sigma$ tension with the CMB. While this difference might not be large enough to cause real concern, it is worthwhile to note that a non-stable DM sector could provide a straight-forward explanation for a reduced clustering signal towards low redshifts \citep{Wang2010, MacCrann2015,Cen2001}. Hence, a decrease of the this tension has indeed been reported for both one-body \citep{DES,Berezhiani:2015yta,Enqvist15,Chudaykin:2016yfk,Archidiacono2019} and two-body decay scenarios \citep{Abellan,Cheng2015,Aoyama2011,Peter2010,Peter2010_2,Wang2014}.

A more significant tension has emerged between the Hubble parameter obtained from supernova distance measurements \citep{Riess:2016jrr,Riess:2019cxk} and derived via the CMB experiment Planck \citep{Planck}. Different one- and two-body decaying DM scenarios with both early and late-time decays have been proposed as a potential solution to the Hubble tension \citep[e.g.][]{Pandey:2019plg,Blinov:2020uvz,Vattis:2019efj}. However, to what extend such models may indeed reconcile CMB with the supernova data \citep{Clark:2020miy,Haridasu:2020xaa} remain disputed. A summary of potential solutions to the Hubble tension, including decaying dark matter (DDM) models, can be found in Ref.~\citep{DiValentino:2021izs}.

In the present paper, we investigate the possibility that part or all of the DM consist of an unstable particle decaying into radiation. The model is fully parametrised via the particle decay rate $\Gamma=\tau^{-1}$ and the fraction of decaying-to-total dark matter, $f=\Omega_{\rm ddm}/\Omega_{\rm dm}$. The radiation particle is generally assumed to be part of the dark sector as well, since decays into photons or other particles of the standard model are strongly constrained. Note, however, that since we only investigate gravitational effects caused by the decay process, we do not need to specify the nature of the radiation species.

In this paper we pursue two main goals. First, we study the effect of dark matter decays on nonlinear structure formation. We modify the $N$-body code {\tt Pkdgrav3} \citep{Potter16} to include decaying DM both at the particle and the cosmological background level. With this at hand, we run a suite of simulations and construct a fitting function describing the suppression effect on the matter power spectrum. As a second step, we perform a forecast analysis for stage-IV weak-lensing (WL) surveys based on a Markov chain Monte Carlo (MCMC) analysis with $\Lambda$CDM mock data and a simulated covariance matrix. In particular, we include baryonic effects (following the procedure outlined in \citep{Schneider2019a,Schneider2019b}), which cause a similar suppression effect on the matter power spectrum and could therefore be degenerate with the signal from decaying dark matter. With this forecast exercise we want to assess to what extent it is possible to go beyond the current limits that come mainly from the Planck CMB survey.

The paper is structured as follows: In Sec.~\ref{sec:DDMModel} we introduce the decaying DM model and its effects at the level of the cosmological background. Sec.~\ref{sec:Nbody} is dedicated to the $N$-body simulations; we describe the modifications to the code, discuss our simulation suite, and present the fitting function for the matter power spectrum. In Sec.~\ref{sec:forecast} we present the forecast analysis with specific focus on the decaying DM parameters before concluding in Sec.~\ref{sec:Conclusions}.

\section{Decaying dark matter model}
\label{sec:DDMModel}
In this section we describe the one-body DDM 
model and specify the free parameters added to $\Lambda$CDM. The model consists of both a stable CDM component and a DDM component that decays into a form of non-interacting dark radiation (DR). Following the notation of Ref.~\citep{Enqvist15}, the DDM and DR components evolve as
\begin{eqnarray}
\label{eq:DDMevolution}
\dot{\rho}_{\rm ddm} + 3\mathcal{H}\rho_{\rm ddm} = -\Gamma a \rho_{\rm ddm},\hspace{0.5cm}
\dot{\rho}_{\rm dr} + 4\mathcal{H}\rho_{\rm dr} = \Gamma a \rho_{\rm ddm},
\end{eqnarray}
where the density evolution of the CDM component is simply given by $\dot{\rho}_{\rm cdm} + 3\mathcal{H}\rho_{\rm cdm} = 0$. Note that we assume conformal time as the variable, with $\mathcal{H}$ being the conformal Hubble parameter. The DDM decay rate $\Gamma$ is considered a free model parameter and consists of the inverse of the decay half-life, i.e. $\tau =\Gamma^{-1}$. The conformal Hubble parameter is obtained via the Friedmann equation
\begin{equation}
\label{eq:Friedmann}
\mathcal{H}^2 = 8\pi G a^2 \sum_i \rho_{i}/3,
\end{equation}
where $\rho_i$ corresponds to all cosmological components, including CDM, DDM and DR described above. Solving the closed system of Eqs.~\eqref{eq:DDMevolution} and \eqref{eq:Friedmann} yields the full cosmological background evolution.

In order to clarify the notations, in particular with respect to the DDM abundance, we follow the definitions of Ref.~\citep{Enqvist15} and introduce the quantity
\begin{equation}
\label{eq:omegatilde}
F_{i}(a) = \frac{\rho_i(a)}{\rho_c} a^{3(1+w_i)}
\end{equation}
where $w_i$ is the equation-of-state parameter of fluid $i$ and $\rho_c$ is the critical density of the universe.  With the functions $F_i(a)$ at hand, we can now define a set of parameters $\widetilde{\Omega}_i \equiv F_i(a=0)$ as well as $\Omega_i \equiv F_i(a=1)$. These quantities describe the abundance of component $i$ at redshift zero. However, while the former ignores all decay processes, the latter takes them into account. This means that $\Omega_{\rm cdm}=\widetilde{\Omega}_{\rm cdm}$, while $\Omega_{\rm ddm}<\widetilde{\Omega}_{\rm ddm}$, and $\Omega_{\rm dr}>\widetilde{\Omega}_{\rm dr}\equiv0$.

A second free parameter of our DDM model is the fraction of decaying-to-total dark matter. It is defined as
\begin{equation}
f =\widetilde{\Omega}_{\rm ddm}/(\Omega_{\rm cdm}+\widetilde{\Omega}_{\rm ddm}).
\end{equation}
The case $f=1$ means that there is no CDM component, i.e. that all of the dark matter is allowed to decay. This scenario would only agree with current data if the lifetime of the DDM particle is significantly longer than the age of the universe. The case of $f\ll1$, on the other hand, describes a model where only a subdominant dark component decays during cosmic history while the bulk of DM remains stable. In this scenario, the decay rate remains largely unconstrained.


\section{Cosmological \texorpdfstring{$N$}{N}-body simulations}
After defining the DDM scenario, we now discuss the effects on nonlinear structure formation and provide details about the modification of the $N$-body code and the simulations performed. At the end of this section, we show the relative effect of DM decays on the matter power spectrum and provide a fitting function for the observed power suppression effect. Note that throughout this section we assume the cosmological parameters $\Omega_m=0.307$, $\Omega_b=0.048$, $10^9 A_s=2.43$, $h_0=0.678$, and $n_s=0.96$ consistent with the Planck CMB+BAO results from 2018 \citep{Planck}.

\label{sec:Nbody}
\subsection{Implementation}
We implement the DDM model in the $N$-body code \texttt{Pkdgrav3} following the method described in Ref.~\citep{Enqvist15}. This means that we decrease the mass of each  simulation particle according to the relation
\begin{equation}
m(t) = m_i\left[(1-r_{\rm dm}f) + r_{\rm dm}fe^{-\Gamma t}\right],\hspace{0.5cm}r_{\rm dm} = \Omega_{\rm dm}/(\Omega_b + \Omega_{\rm dm}),
\end{equation}
where $m_i$ is the initial particle mass before the decay starts. Note that this equation coincides with Eq. (3.1) of Ref. \citep{Enqvist15}, except that we allow for the additional variation of $f$.

In \texttt{Pkdgrav3}, initial conditions can be generated through the so-called \emph{back-scaling} of a linear transfer function given at $z=0$. This grants sub-percent precision in the linear evolution as well as enabling second-order Lagrangian perturbation theory (2LPT) 
initial conditions generation. In a DDM scenario the \emph{back-scaling} approach is known to produce additional terms when compared to the $\Lambda$CDM prescription \citep{Fidler17}. In principle, it would be possible to account for these terms and keep the \emph{back-scaling} approach, however we opted for modifying the code in a way such that initial conditions are set directly from a given linear matter power spectrum at the starting redshift of the simulation (throughout this work $z_{i}=49$). We use \texttt{CLASS} \citep{CLASSII} to compute the linear matter power spectrum starting from an early redshift $z_i$. Furthermore, we use first-order Lagrangian perturbation theory (1LPT) 
initial conditions generation. 
This approach can be referred to as the \emph{forward} approach and provides a correct prescription for using $N$-body simulations to simulate non-$\Lambda$CDM scenarios.

\subsection{Simulation results}
Based on the modified code described above, we run a suite of $N$-body DDM and CDM simulations varying the two DDM parameters $\tau$ and $f$ assuming $\tau=\{31.6,\,100,\,316\}$ Gyr for the decay lifetime, and $f=\{0.0,\, 0.2,\,0.4,\,0.6,\,0.8,\,1.0\}$ for the decaying-to-total DM fraction. Note that $f=0$ corresponds to the standard $\Lambda$CDM case while $f=1$ means that all DM is allowed to decay.
 
For each model, we use a box size of $L=500$ Mpc/h and a total particle number of $N=1024^3$. Some selected simulations are also run at lower particle numbers of $N=256^3$, $512^3$ in order to test the convergence of our results. 
The initial conditions of the simulations are set up with {\tt Pkdgrav3} at $z_i$, using 
1LPT \citep{Reed2013}. All runs are evolved to redshift zero, producing both particle output boxes plus power spectrum measurements with the {\tt Pkdgrav3} internal on-the-fly power spectrum calculator. A list of all simulations is provided in Table~\ref{tab:simlist}.

\begin{table}[tb]
    \centering
    \begin{tabular}{c c c c}
        $N$ & $L$ [Mpc/$h$] & $\tau$ [Gyr] & $f$ \\
        \hline
        1024$^3$ & 500 & 31.6 & 1, 0.8, 0.6, 0.4, 0.2, 0 \\
        1024$^3$ & 500 & 100 & 1, 0.8, 0.6, 0.4, 0.2, 0 \\
        1024$^3$ & 500 & 316 & 1, 0.4, 0\\
        512$^3$ & 500 & 100 & 1, 0.3, 0\\
        256$^3$ & 500 & 316 & 1, 0\\
        \hline
    \end{tabular}
    \caption{List of $N$-body simulations presented in this work. $N$ is the particle number, $L$ is the simulation box size, $\tau=1/\Gamma$ the decay lifetime, and $f$ the decaying-to-total DM fraction. Note that $f=1$ means that all DM is allowed to decay while $f=0$ refers to the $\Lambda CDM$ simulation.}
    \label{tab:simlist}
\end{table}

\begin{figure}[tb]
\centering 
\includegraphics[width=\textwidth, trim=0.2cm 0.3cm 0.3cm 0.3cm, clip]{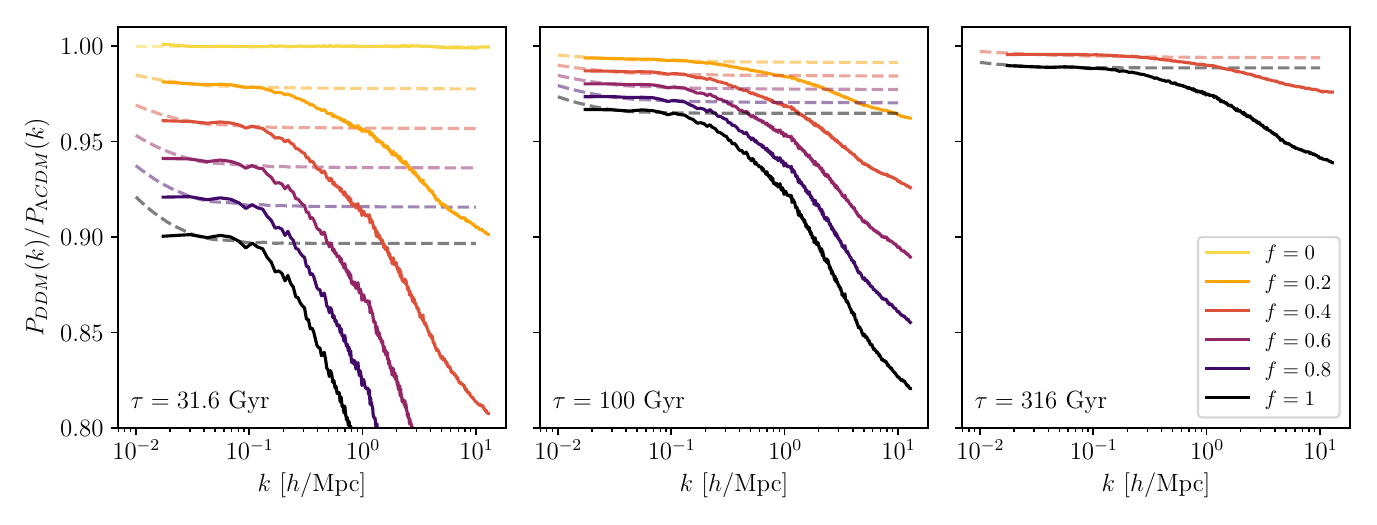}
\caption{$P_{DDM}(k)/P_{\Lambda CDM}(k)$ for the N-body simulations ran with \texttt{Pkdgrav3} at $z=0$ and a varying fraction $f$ (solid lines). We also show the same ratio for \texttt{CLASS} runs, where the same DDM parameters have been used (dashed lines).}
\label{fig:1_body_results}
\end{figure}

In Fig.~\ref{fig:1_body_results} we plot the ratio of the DDM power spectra divided by the standard CDM case at redshift zero. The left-hand, central, and right-hand panels correspond to three different decay lifetimes while the coloured solid lines represent simulation results assuming different decaying-to-total DM fractions. The dashed lines show the linear results obtained with {\tt CLASS}.

The resulting power suppression of DDM relative to the standard CDM model is of non-trivial shape. A first smooth step-like suppression can be observed in the linear regime at around $k\sim10^{-2}$ h/Mpc. While this feature is visible in the results from {\tt CLASS}, we do not resolve it with our simulations due to the relatively small box-size of 500 Mpc/h. After a visible plateau at $k\sim 0.2-2$ h/Mpc, the ratio between DDM and CDM power spectra decreases again. This second suppression is an effect caused by the nonlinear behaviour of structure formation. It is therefore only visible in the simulation results but not in the linear results from {\tt CLASS}.

A second notable effect visible in Fig.~\ref{fig:1_body_results} is that both DDM parameters ($\tau$, $f$) have a direct influence on the overall amplitude of the suppression but not so much on the scales of the first and second step-like suppression feature. These steps manifest at $k\sim10^{-2}$ and $\sim 10^{-1}$ h/Mpc, independently of the values assumed for $\tau$ and $f$. Similar conclusions have been drawn by the authors of Ref.\citep{Enqvist15}. Note that this particular behaviour makes it possible to come up with a general fitting function for the DDM power suppression as we will discuss in Sec.~\ref{sec:Fitting1body}.

We investigate the convergence properties as well as comparing our code to published results from Ref.~\citep{concept} in Appendix~\ref{sec:AConcept}. Summarizing, the analysis of the simulation convergence together with the cross-check with the {\tt CONCEPT} \citep{concept} $N$-body code lets us conclude that our results are reliable down to at least $k\approx$ 6.4 h/Mpc (and possibly up to somewhat higher values).

\subsection{Fitting function}
\label{sec:Fitting1body}
In this section we present an updated fitting function for the DDM to CDM power ratio of the DDM model with free decay lifetime $\tau$ and decaying-to-total DM fraction $f$. The function is inspired by the fit of Ref.~\citep{Enqvist15}, however, with the fraction $f$ as an additional parameter. We also re-calibrate the fit to our results. 
We can write the nonlinear suppression as $P_{\rm DDM}(k)/P_{\rm \Lambda CDM}(k) = 1 - \varepsilon_{\rm nonlin}(k)$, with
\begin{equation}\label{fit}
\frac{\varepsilon_{\rm nonlin}(k)}{\varepsilon_{\rm lin}} = \frac{1+a(k/\text{Mpc}^{-1})^p}{1+b(k/\text{Mpc}^{-1})^q}f,
\end{equation}
where the factors $a$, $b$, $p$, and $q$ are given by
\begin{eqnarray}
a(\tau,z) &=& 0.7208 + 2.027\left(\frac{\text{Gyr}}{\tau}\right) + 3.031\left(\frac{1}{1+1.1z}\right) - 0.18,\nonumber\\
b(\tau,z) &=& 0.0120 + 2.786\left(\frac{\text{Gyr}}{\tau}\right) + 0.6699\left(\frac{1}{1+1.1z}\right) - 0.09,\nonumber\\
p(\tau,z) &=& 1.045 + 1.225\left(\frac{\text{Gyr}}{\tau}\right) + 0.2207\left(\frac{1}{1+1.1z}\right) - 0.099,\nonumber\\
q(\tau,z) &=& 0.9922 + 1.735\left(\frac{\text{Gyr}}{\tau}\right) + 0.2154\left(\frac{1}{1+1.1z}\right) - 0.056.\nonumber
\end{eqnarray}
The function $\varepsilon_{\rm lin}$ describing the main redshift evolution of the signal is given by 
\begin{equation}
\varepsilon_{\rm lin}(\tau, z) = \alpha\left(\frac{\rm Gyr}{\tau}\right)^{\beta}\left(\frac{1}{(0.105z)+1}\right)^{\gamma},
\end{equation}
where $\alpha$, $\beta$, $\gamma$ are functions of $\omega_b$, $h_{\rm  ddm}$, and $\omega_m = \omega_b + \omega_{\rm  dm}$ given by
\begin{eqnarray}
\alpha & = & (5.323 - 1.4644u - 1.391v) + (-2.055 + 1.329u + 0.8672v)w +\nonumber\\
& & (0.2682 - 0.3509u)w^2,\nonumber\\
\beta & = & 0.9260 + (0.05735 - 0.02690v)w + (-0.01373 + 0.006713v)w^2,\nonumber\\
\gamma & = & (9.553 - 0.7860v) + (0.4884 + 0.1754v)w + (-0.2512 + 0.07558v)w^2,\nonumber
\end{eqnarray}
with $u=\omega_b/0.02216$, $v=h_{\rm ddm}/0.6776$ and $w = \omega_m/0.14116$. The fitting function is able to reproduce our simulation results by better than one percent up to $k=13$ h/Mpc and for redshifts between 0 and 2.35. This is significantly better than the 10 percent accuracy reported by Ref.~\citep{Enqvist15} who only focused on the case of $f=1$.

\begin{figure}[tb]
\centering 
\includegraphics[width=\textwidth, trim=0.4cm 0.3cm 0.3cm 0.3cm, clip]{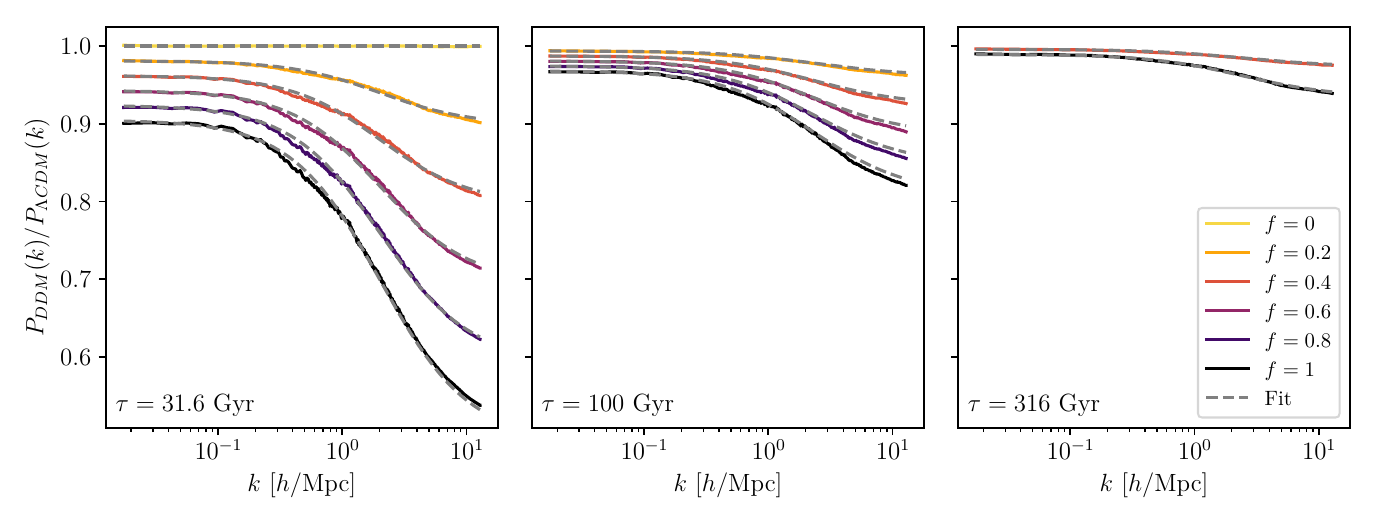}
\includegraphics[width=\textwidth, trim=0.4cm 0.3cm 0.3cm 0.3cm, clip]{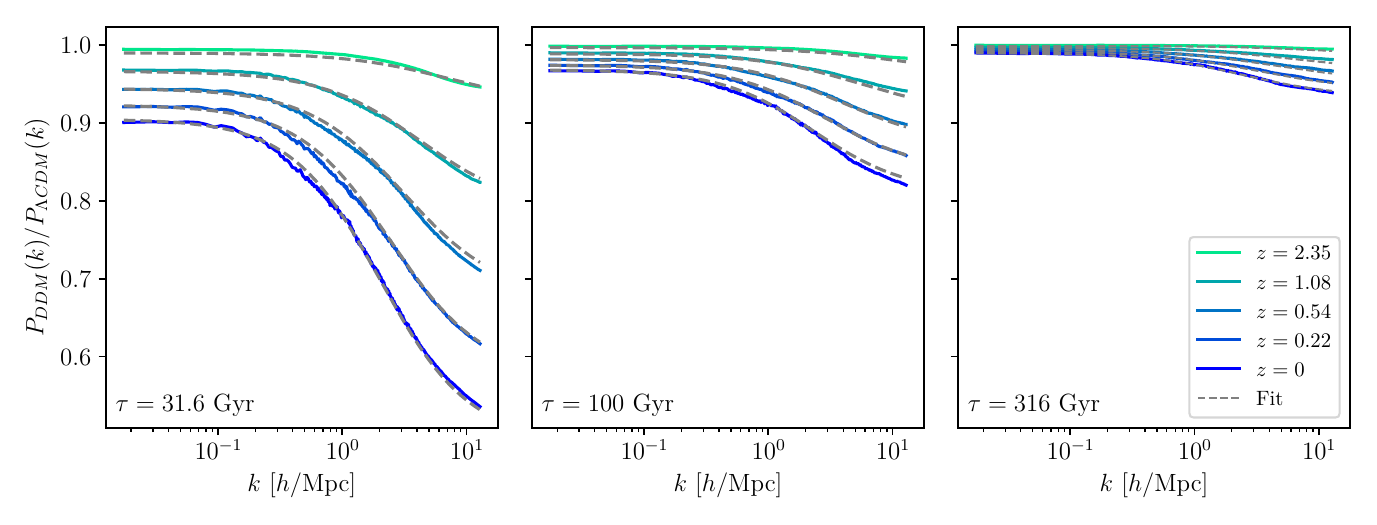}
\caption{Matter power spectrum ratio $P_{DDM}(k)/P_{\Lambda CDM}(k)$ for N-body simulations run with \texttt{Pkdgrav3}. Top row: simulation results at $z=0$ and with a varying fraction $f$ parameter for the decay lifetimes $\tau=31.6 \ \text{Gyr}, \ 100 \ \text{Gyr},\ 316 \ \text{Gyr}$ (solid lines). Bottom row: simulation results at $z=0,\ 0.22,\ 0.54,\ 1.08,\ 2.35$ with a fixed fraction $f=1$ parameter for the decay lifetimes $\tau=31.6 \ \text{Gyr}, \ 100 \ \text{Gyr},\  316 \ \text{Gyr}$ (solid lines). Overlapped we show the fitting formula described in Sec.\ref{sec:Fitting1body} (dashed lines).}
\label{fig:fit_z0}
\end{figure}

In Fig.~\ref{fig:fit_z0} we compare the fitting function to our simulation results, assuming three different decay lifetimes $\tau=31.6$, 100, and 316 Gyr (left to right). The top-panels refer to the $z=0$ case with varying decaying-to-total DM fraction ($f$); the bottom-panels focus on the $f=1$ case showing the redshift evolution from $z=0$ to 2.35. Note that a similar agreement is obtained for the redshift evolution of the other models with decaying-to-total DM fraction below one.

The fit of Eq.~\ref{fit} has been calibrated to simulations based on the Planck cosmology. However, it is argued in Ref.~\citep{Enqvist15} that varying the cosmological parameters should only marginally affect the validity of the fit as long as the parameters do not significantly deviate from the fiducial ones.

\section{Parameter forecast analysis}
\label{sec:forecast}
After having quantified the nonlinear effects of decaying DM on the power spectrum, we will now investigate how well DM decays can be constrained with future galaxy surveys. We specifically focus on the tomographic weak-lensing shear power spectrum as a potential observable, since weak lensing probes the medium to small cosmological scales (up to $k\sim 10$ h/Mpc) where the effect from DDM is strongest.

Our forecast study follows the analysis presented in Refs.~\citep{Schneider2019a,Schneider2019b}. It is based on a realistic Euclid-like mock data set with three tomographic bins between redshift 0.1 and 1.5 and with a simulation-based covariance matrix from statistically independent light-cone realisations that include a simplified survey footprint, shape-noise errors, and sample variance. The mock data is based on a $\Lambda$CDM universe and accounts for baryonic effects via the \emph{baryonification} model of Refs. \citep{SchneiderTeyssier2015, Schneider2018}. This is particularly important since baryonic feedback leads to a suppression effect of the power spectrum that is qualitatively similar to the effect from DDM \citep[see e.g. Refs.][]{vanDaalen2019,Chisari2019}. In this study we consider the baryonification model that depends on three parameters ($M_c$, $\mu$, $\theta_{ej}$). These parameters model the gas profile in haloes of mass $M$, which has the following form:
\begin{eqnarray}
\rho_\mathrm{gas}(r) &\propto \frac{1}{\left[1+ 10(r/r_\mathrm{v}) \right]^{3-(M_c/M)^\mu} \left[1+ r^2/(\theta_\mathrm{ej}r_\mathrm{v})^2 \right]^{2+0.5(M_c/M)^\mu}} \ , 
\end{eqnarray}
where $r_\mathrm{v}$ is the virial radius (defied at 200 times the critical density). The same three-parameter baryonification model has been used to recover the power spectrum suppression seen in hydro-dynamical simulations \cite[for a detailed comparison, see Ref.][]{Schneider2018}.

The prediction of the weak-lensing shear power spectrum is based on the Limber approximation and the halofit matter power spectrum from Refs.~\citep{Takahashi2012,Bird}. The baryonic effects are included using the emulator of Ref.~\citep{Schneider2019a}. Note that this emulator has been built assuming a $\Lambda$CDM cosmology. Although it is, strictly speaking, not guaranteed to provide exact results for decaying DM scenarios as well, we do not expect it to introduce substantial biases. Our expectation is based on recent results from Ref.~\cite[][]{parimbelli2021mixed}, who showed nearly perfect separability between baryonic effects and other non-minimal dark matter scenarios. We therefore defer to  future work for a more systematic analysis of baryonic effects on DDM simulations.

The power suppression due to DM decays is described via the the fitting function presented in Sec.~\ref{sec:Fitting1body}. Note that the absence of any dependence on baryonic parameters in the fit reflects the perfect separability of DDM and baryonic effects discussed above.

\begin{figure}[tb]
\centering 
\includegraphics[width=0.495\textwidth, trim=0.5cm 0.3cm 1.5cm 1.1cm, clip]{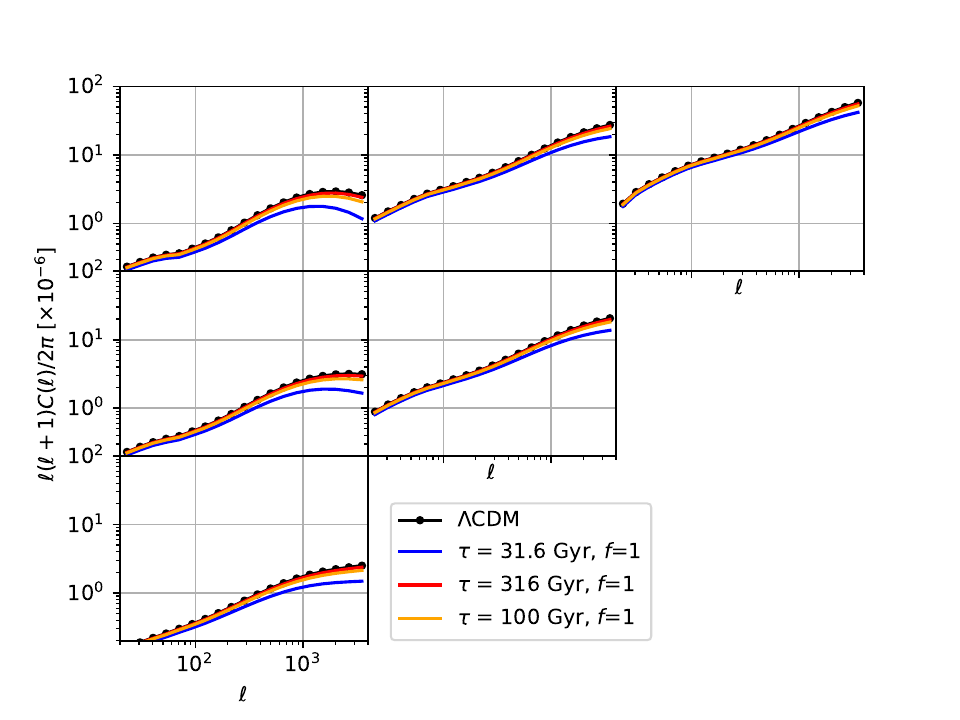}
\includegraphics[width=0.495\textwidth, trim=0.5cm 0.3cm 1.5cm 1.1cm, clip]{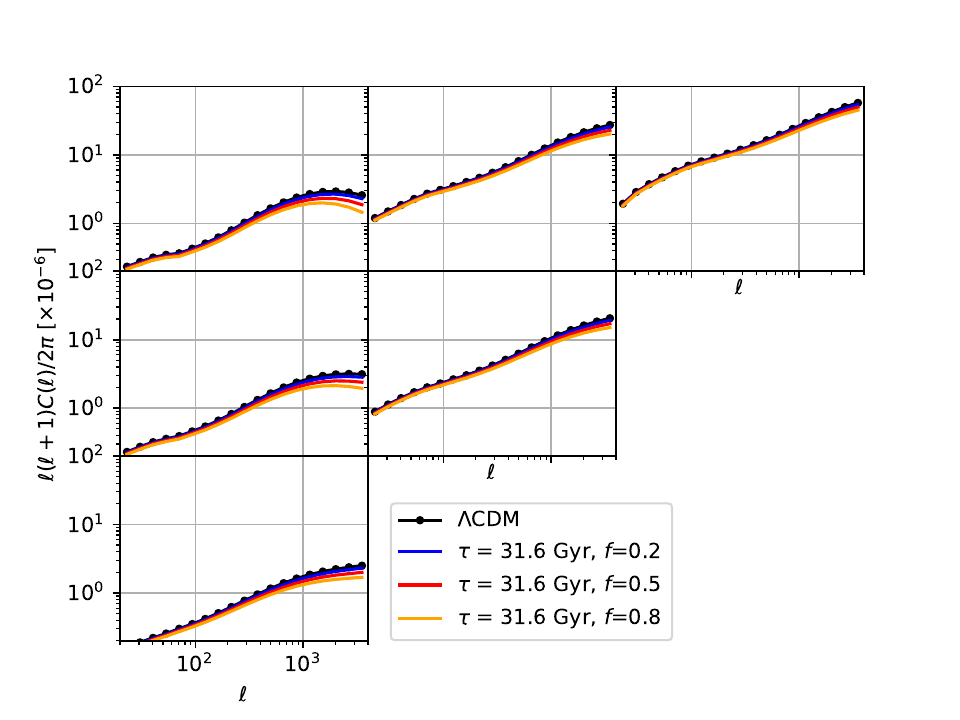}
\caption{
Mock auto and cross shear power spectra expected from a Euclid-like survey for various DDM scenarios. For comparison, the shear power spectra corresponding to the CDM universe is shown as lines with circular markers.
We assume three redshift bins with edges at $z=0.1, 0.478, 0.785$ and $1.5$.
In the left- and right-hand panels, we vary decay lifetime and decay fraction, respectively.}
\label{fig:Cls_DDM}
\end{figure}

In Fig.~\ref{fig:Cls_DDM} we illustrate the impact of DDM models on the tomographic angular power spectrum of cosmic shear. While the black data points represent our Euclid-like mock data based on pure $\Lambda$CDM, the coloured lines show predictions for various DDM models. In the left-hand panel, we keep the DDM fraction fixed to $f=1$ and only vary the decay lifetime $\tau$. In the right-hand panel, we fix the DDM lifetime to $\tau=31.6$ Gyr and vary the DDM fraction $f$. As expected from the shape of the matter power spectrum, the effects from dark matter decays are visible over all scales but are most pronounced towards the highest $\ell$-values.

Based on the Euclid-like mock data and the prediction-pipeline discussed above, we perform an MCMC likelihood analysis to estimate the constraining power for one-body DDM. We thereby investigate two scenarios, one where the neutrinos are assumed to be mass-less (Sec.~\ref{sec:ForecastMassless}) and one where we vary (and marginalise over) neutrino masses between 0.06 and 0.8 eV (Sec.~\ref{sec:ForecastMassive}). While we know that neutrinos have non zero mass, the first scenario remains interesting, as it allows for a direct comparison with previous results from the literature. Next to the sum of the neutrino masses ($\Sigma m_{\nu}$) and the two DDM parameters ($\tau$, $f$), we furthermore vary five cosmological parameters ($\Omega_m$, $\Omega_b$, $A_s$, $n_s$, $h_0$), one intrinsic alignment parameter ($A_{\text{IA}}$, based on the model of Ref.~\citep{Hirata&Seljak}), plus three baryonic parameters ($M_c$, $\mu$, $\theta_{ej}$) from the \emph{baryonification} model of Ref.~\citep{Schneider2018}. This means that most our MCMC runs are based on a total of 11 (mass-less neutrino case) and 12 (varying neutrino mass case) free parameters, respectively.

For the intrinsic alignment and baryonic parameters, we use flat priors that are wide enough to not affect the obtained posterior results. One notable exception is the parameter $\mu$ that remains largely unconstrained by our mock data. Note, however, that in any case, the baryonic priors are selected to be wide enough so conservatively include a large variety of different results predicted by hydrodynamical simulations ~\citep[see Ref.][for more details]{Schneider2019a}. The prior ranges are listed in table~\ref{tab:priorranges}.

Regarding the cosmological parameters, we first perform a weak-lensing-only analysis with flat, uninformative priors. This corresponds to the case where no other information other than the one from the Euclid-type WL mock is included. In a second step, we also include Gaussian priors from the Planck CMB experiment assuming the latest data release \citep{Planck}. 

In this context, it is important to notice that the CMB data is not only able to constrain the standard cosmological parameters, but it also provides constraints on the DDM parameters themselves. This may be surprising at first, since recombination happens well before the start of any DM decays (at least for the models investigated here). However, since the CMB contains information about the late-time cosmic growth history via the Sachs-Wolfe effect, it is also sensitive to the DDM parameters as they lead to evolving $\Omega_m$ and $\Omega_r$. We obtain CMB priors for the DDM parameters using results from Ref.~\citep{Poulin16}. The details about how these priors are extracted are outlined in Appendix \ref{sec:APoulin}.

\begin{table}[]
    \centering
    \small
    \begin{tabular}{c c c c c}
         Parameters & Mock value & WL priors & WL+CMB priors & WL+CMB+fixed baryons priors \\
         \hline \hline
          $\Omega_b$ & 0.049 & 0.04 - 0.06 & 0.001 & 0.001 \\
          $\Omega_m$ & 0.315 & 0.25 - 0.42 & 0.0084 & 0.084\\
          $10^9A_s$ & 2.035 & 1.4 - 3.5 & 0.034 & 0.034\\
          $n_s$ & 0.966 & 0.9 - 1.0 & 0.0044 & 0.0044\\
          $h_0$ & 0.673 & 0.6 - 0.9 & 0.006 & 0.006\\
          \hline
          $\Sigma m_{\nu}$ (Sec.~\ref{sec:ForecastMassless})& 0 &  0& 0 & 0\\
          $\Sigma m_{\nu}$ (Sec.~\ref{sec:ForecastMassive}) & 0.1 & 0.06 - 0.8 & 0.1 & 0.1\\
          \hline
          $A_{IA}$ & 1 & 0 - 2 & 0 - 2 & 0 - 2\\
          \hline
          $\log M_c$ & 13.8 & 13 - 16 & 13 - 16 & 0\\
          $\mu$ & 0.21 & 0.1 - 0.7 & 0.1-0.7 & 0\\
          $\theta_{ej}$ & 4 & 2 - 8 & 2 - 8 & 0\\
         \hline
         $\Gamma$ & 0 & 10$^{-6}$ - 0.0316 & see below & see below\\
         $f$ & 0 & 0.0 - 1.0 & see below & see below\\
         \hline
         $\Gamma f^{c_3}$ & 0 & not used & $(c_2 + c_1 f^{c_3})/2$ & $(c_2 + c_1 f^{c_3})/2$\\
         \hline
    \end{tabular}
    \caption{Priors of cosmological, neutrino, intrinsic-alignment ($A_{IA}$), baryonic parameters, and DDM parameters for the WL, WL+CMB, and WL+CMB+fixed baryons analysis. Whenever we assume a flat prior, we specify the range (lower limit - upper limit). For Gaussian priors, we provide just one number (standard deviation). For the DDM Gaussian priors we follow the approach outlined in Appendix \ref{sec:APoulin}, using $c_1=0.00721714,\ c_2=0.00308819,\ c_3=1.45599761$. Note that neutrino masses are assumed fixed and massless in Sec.~\ref{sec:ForecastMassless} and only varied in Sec.~\ref{sec:ForecastMassive}.}
    \label{tab:priorranges}
\end{table}


\subsection{Forecast assuming massless neutrinos}
\label{sec:ForecastMassless}
As a first step, we perform an MCMC analysis with the sum of the neutrino masses fixed at zero ($\Sigma m_{\nu}=0$). While this assumption is known to be inadequate, it considerably simplifies the analysis and allows us to directly compare to previous work on decaying DM \citep[which did not vary neutrino masses, see e.g. Refs.][]{Enqvist15,Poulin16,Enqvist19}.

\begin{figure}[tbp]
\centering
\includegraphics[width=0.423\textwidth,trim=0.4cm 0.7cm 1.0cm -0.0cm,clip]{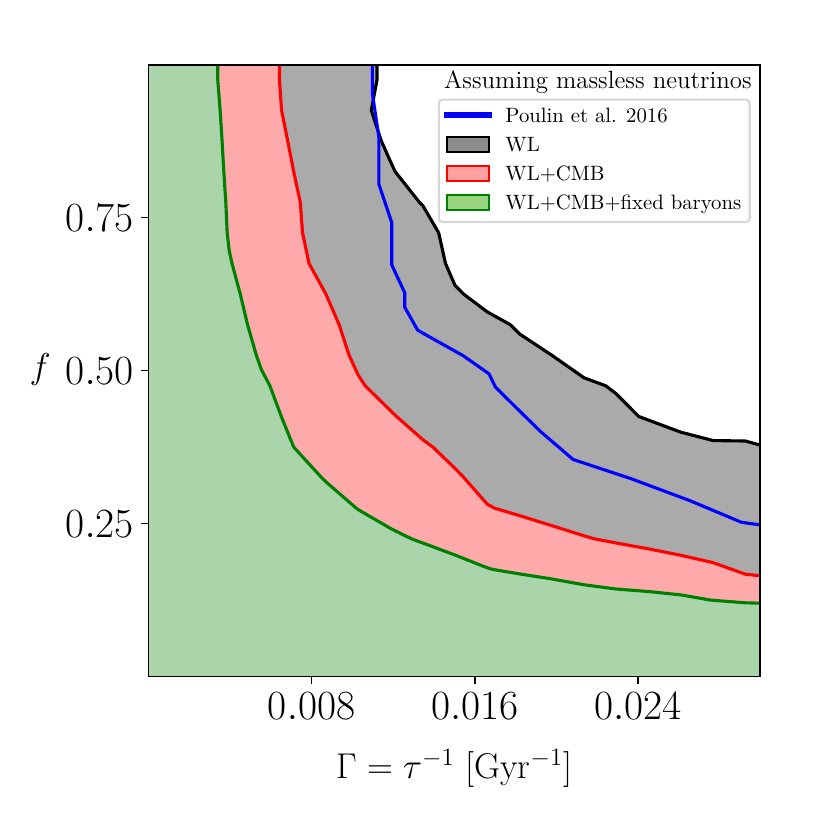}
\includegraphics[width=0.55\textwidth,trim=0.15cm -0.2cm 1.4cm 1.3cm,clip]{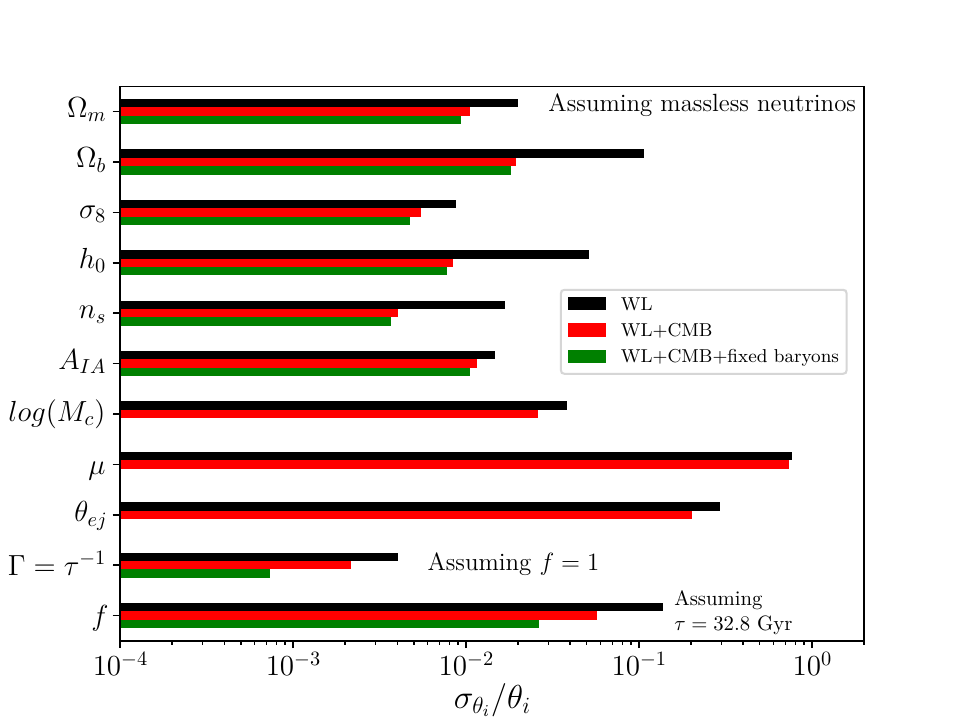}
\caption{Results for our MCMC analysis assuming massless neutrinos. In black we show results when assuming exclusively WL observables ({$WL$}), in red we additionally assume CMB priors ({$WL$+CMB}) and in green we additionally fix the baryonic parameters $log(M_c), \ \mu, \ \theta_{ej}$ to their mock value. We report results from Ref.\citep{Poulin16} (Poulin et al. 2016) in blue. Contours in the left panel are reported at 95 percent CL In the right-hand panel we show marginalised errors ($\sigma_{\theta_i}$) for all the parameters, normalized by their mock value ($\theta_i$), reported on 68 percent CL}
\label{fig:Results_nomnu}
\end{figure}

We start by performing an MCMC run only considering the WL mock data and assuming flat, uninformative priors (as specified in the third column of table~\ref{tab:priorranges}). This corresponds to the hypothetical case where Euclid-like WL data is used as an independent probe with no added information --- neither from galaxy clustering nor from any external data source.

The resulting constraints on the DDM parameters for the WL-only case are shown as grey contours in the left-hand panel of Fig.~\ref{fig:Results_nomnu}. Comparing it to the current constraints based on the Planck data from Ref.~\citep{Poulin16} (shown as blue line), we conclude that the Euclid-like WL data will have a similar constraining power than Planck for the $f=1$ case, while it will lead to weaker constraints for $f<1$.

As a further step, we run an MCMC analysis combining Euclid-like weak-lensing data with CMB data from Planck. We do this by adding Gaussian priors from the Planck18 data set to both the cosmological and the DDM parameters. The priors of the DDM parameters are obtained from Ref.~\citep{Poulin16} following the method outlined in Appendix \ref{sec:APoulin}. Detailed information about the width of the Gaussian priors can be found in the fourth column of table.~\ref{tab:priorranges}. The forecast posteriors of the combined WL+CMB analysis is shown as red contours in the left-hand panel of Fig.~\ref{fig:Results_nomnu}. The contours are about 50 percent tighter than the current constraints from Ref.~\citep{Poulin16} (blue line) for all values of $f$.

So far, we have marginalised over baryonic parameters assuming that we have no prior knowledge of the baryonic suppression effect on the matter power spectrum. In Ref.~\citep{Schneider2019b}, however, it was shown that future X-ray data may be used to strongly constrain the baryonic parameters, e.g. via the observed fraction of gas in haloes. In order to estimate how much can be gained by including additional information about baryons, we run another MCMC chain fixing the baryonic parameters to their default values.

The resulting posteriors of the WL+CMB analysis with fixed baryons is shown as green contours in the left-hand panel of Fig.~\ref{fig:Results_nomnu}. They are 40 percent tighter compared to the case with free baryonic parameters and 60 percent tighter compared to current constraints from Planck. As there is a noticeable degeneracy between the DDM parameters and the baryonic feedback parameters, fixing the latter leads to a significant improvement in constraining power. It highlights the importance of better understanding and constraining baryonic feedback for future WL surveys. In  Appendix~\ref{sec:AForecast} we show the complete posterior distributions from our MCMC runs. More details about parameter degeneracies can be found there.

Regarding the special case of $f=1$ (i.e. all DM being allowed to decay), we obtain $\tau\geq 89.08$ Gyr for the WL only case, $\tau\geq 154.19$ Gyr for the WL+CMB case, and $\tau\geq 289.28$ Gyr for the WL+CMB case with fixed baryonic parameters. For the limit of very large decay rates ($\tau<34$ Gyr), on the other hand, we obtain limits on the decaying-to-total DM fraction of $f<0.38$ for the WL-only case, $f<0.17$ for the WL+CMB case, and $f<0.12$ for WL+CMB case with fixed baryonic parameters. Note that all these limits are at 95 percent CL.

Ref.\citep{Poulin16} reports a lower bound of 3.8 percent for the fraction of DDM that has decayed between recombination and today (at 95 percent CL). We emphasize that this is not the same constraint that we obtain on $f$ when observing the limit of small decay lifetimes (our smallest bin being $31.6 $ Gyr $< \tau < 34$ Gyr). In our case the fraction $f$ represents the decaying-to-total DM fraction at the beginning of the simulation. Tighter constraints on $f$ could be achieved by exploring cases with small decay lifetimes (smaller than the age of the Universe), a domain we do not explore in this work.

The right-hand panel of Fig.~\ref{fig:Results_nomnu} summarises the marginalised constraints (now at 68 percent CL) of all parameters in our MCMC forecast analysis. It shows that adding prior information from the CMB significantly decreases the uncertainties of all cosmological and DDM parameters. Unsurprisingly, the largest improvements are obtained for the cosmological parameters $h_0$ and $\Omega_b$, which are known to be rather poorly constrained by WL data. Additionally fixing baryonic parameters only has a relatively mild effect (of order 20 percent) on the cosmological parameters ~\citep[as pointed out in Refs.][]{Schneider2019a,Schneider2019b}. However, it significantly affects the two DDM parameters whose 68 percent CL constraints become tighter by more than a factor of 2. This strong improvement is a result of the fact that baryonic and DDM parameters affect the matter power spectrum in similar ways and are therefore inherently degenerate. 

The results presented in this section indicate that with upcoming WL surveys such as Euclid or LSST, it will be possible to go significantly beyond current constraints, probing unexplored parts of the DDM parameter space. 
In the next section we will investigate to what extent this conclusion still holds if we vary the neutrino masses as an additional free model parameter.

\subsection{Forecast with varying neutrino masses}\label{sec:ForecastMassive}
Upcoming weak-lensing and galaxy clustering surveys will be sensitive enough to probe and constrain the sum of the neutrino masses \citep[see e.g. Ref.][]{Amendola:2016saw}. We therefore repeat the analysis above, now varying and marginalising over the $\Sigma m_{\nu}$ parameter. While this is a more consistent approach, it is worth noticing that the results cannot be directly compared to previous limits that were obtained assuming neutrinos to be mass-less \citep[e.g.][]{Enqvist15,Enqvist19,Poulin16,DES}.

\begin{figure}[tbp]
\centering
\includegraphics[width=0.423\textwidth,trim=0.4cm 0.7cm 1.0cm 0.0cm,clip]{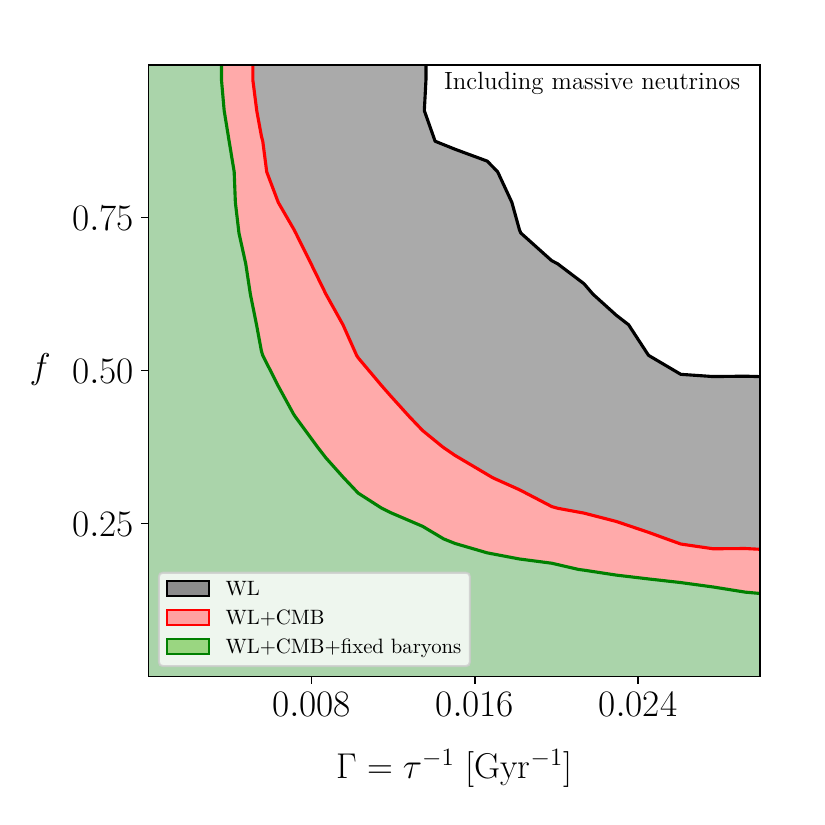}
\includegraphics[width=0.55\textwidth,trim=0.15cm -0.2cm 1.4cm 1.3cm,clip]{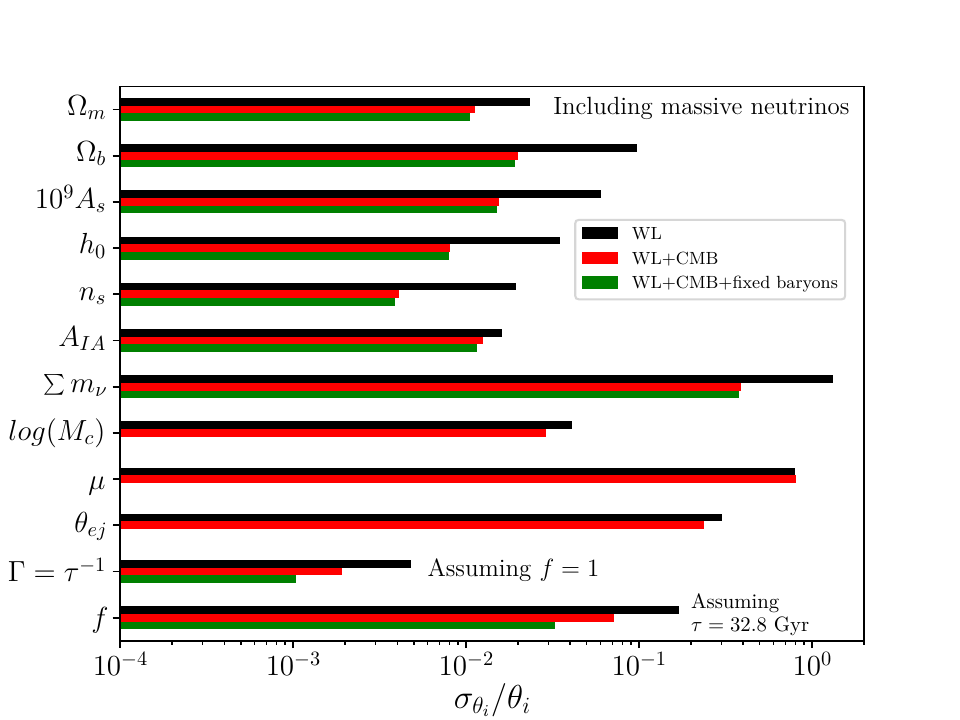}

\caption{Results for the MCMC analysis with varying neutrino masses. The weak-lensing  only case (WL) is shown in grey, the scenarios with additional CMB priors (WL+CMB) and with CMB priors plus fixed baryonic parameters (WL+CMB+fixed baryons) are shown in red and green. The CMB priors are obtained from the Planck18 analysis \citep{Planck}. The left-hand panel shows the posteriors of the decaying DM parameters (at 95 percent CL) with all other parameters marginalised over. The right-hand panel provides an overview over all marginalised errors ($\sigma_{\theta_i}$) at 68 percent CL normalized to their mock values ($\theta_i$). For the special cases of $\tau$ and $f$ (where the mock values are zero), we show the absolute errors at the limits $f=1$ and $\tau=32.8$ Gyr, instead.}
\label{fig:Results_mnu}
\end{figure}

Since the true value of $\Sigma m_{\nu}$ is not known, we first have to select a reasonable mass for our mock data. We assume $\Sigma m_{\nu}=0.1$ eV based on a normal mass hierarchy, a setup that is well within the current neutrino mass constraints \citep{Planck,Palanque-Delabrouille:2015pga}. More details about our mock data for a $\Lambda$CDM case with massive neutrinos can be found in Ref.~\citep{Schneider2019b}.


The performed MCMC runs are based on the same setup as in the previous section. We first assume a WL-only case before adding priors for the cosmological, DDM, and neutrino parameters from the Planck18 CMB experiment (WL+CMB case). Finally, we investigate a third scenario where baryonic parameters are fixed, mimicking the ideal case of strongly constrained baryonic feedback from e.g. X-ray observations \citep{Schneider2019a,Schneider2019b}.

In the left-hand panel of Fig.~\ref{fig:Results_mnu} we plot the posteriors of the decaying DM parameters including varying neutrino masses. The contours of the three scenarios are again shown in grey (WL-only), red (WL+CMB), and green (WL+CMB with fixed baryonic parameters) at 95 percent CL. Compared to the results with fixed neutrino masses (see Fig.~\ref{fig:Results_nomnu}), the contours of the WL only case are somewhat looser, while the WL+CMB cases (without and with fixed baryonic parameters) remain very similar. Note that although we generally expect limits to become weaker when adding a new model parameter, this does not have to be the case here since we have also changed the mock data (now assuming a $\Lambda$CDM model with $\Sigma m_{\nu}=0.1$ eV).

Regarding the special case of $f=1$, we obtain constraints on the decay lifetime of $\tau \geq 74.88$ Gyr for the case of WL only, $\tau \geq 182.23$ Gyr when additionally introducing CMB priors and $\tau \geq 275$ Gyr when the baryonic parameters are fixed. Considering the limit of large decay rates and small decaying-to-total DM fraction, we obtain a maximum allowed fraction of $f<0.49$ for the WL only case, $f<0.21$ for WL+CMB case, and $f<0.13$ for the WL+CMB case with fixed baryonic parameters.

In the right-hand panel of Fig.~\ref{fig:Results_mnu} we show the marginalised errors (at 68 percent CL) of all cosmological, intrinsic alignment, baryonic, and DDM parameters. As in the previous case, we notice that the errors on $h_0$ and $\Omega_b$ improve most when combining WL with CMB priors. We also confirm that by fixing the baryonic parameters, the constraints on the DDM parameters improve by about another factor of 2. The constraint on the neutrino parameter ($\Sigma m_{\nu}$) on the other hand, only slightly improves when fixing the baryonic parameters. This suggests that neutrino and baryonic parameters are not degenerate, a fact that has been pointed out previously \citep{Mummery:2017lcn,Parimbelli:2018yzv,Schneider2019b}. 

Overall, we confirm that the general conclusions obtained with the assumption of mass-less neutrinos also hold for the more general case of massive neutrinos with varying $\Sigma m_{\nu}$. Although the constraints on the DDM parameters become  loose for the WL only case, this difference shrinks substantially once CMB priors are included. For the WL+CMB cases, fixing or marginalising over the neutrino mass leads to very similar DDM constraints. We therefore conclude that even in the realistic case of a $\Lambda$CDM universe with massive neutrinos, the the one-body DDM model can be efficiently constrained by combining upcoming WL observations with data from the CMB.

\section{Conclusions}
\label{sec:Conclusions}
In this paper we have investigated the nonlinear effects on the matter power spectrum, assuming a decaying dark matter scenario where parts or all of the dark matter may decay into (dark) radiation. We have modified the $N$-body code {\tt Pkdgrav3}, \citep[following an approach described in Ref.~][]{Enqvist15} and run a suite of simulations with varying decay lifetime ($\tau$) and decaying-to-total DM fraction ($f$). Based on these simulations, we defined a fitting function for the relative effect of decaying DM on the matter power spectrum. The fit is sub-percent accurate up to $k= 12$ h/Mpc, making it a promising tool to probe DM decays with future weak-lensing and galaxy-clustering surveys. 

With the fit of the nonlinear power spectrum from simulations, we performed a forecast analysis for a Euclid-like tomographic weak-lensing survey using $\Lambda$CDM mock data and a simulation-based covariance matrix \citep[following the method outlined in Refs.~][]{Schneider2019a,Schneider2019b}. We first ran an MCMC analysis, simultaneously varying cosmological, intrinsic alignment, baryonic, and decaying DM parameters, while assuming neutrinos to be mass-less. With this setup in hand, we showed that, while Euclid-like weak-lensing data with flat, uninformative priors will lead to weaker limits on decaying DM compared to current constraints from the CMB \citep{Poulin16}, a combination of WL with CMB data will make it possible to significantly push into  currently unexplored DDM parameters space. This is even more the case for the optimistic scenario where baryonic feedback effects can be fully determined using external data \citep[e.g. from the X-ray survey eROSITA, see Ref.][]{Schneider2019b}. In this case we expect an improvement on the DDM parameter constraints (for $\tau$ and $f$) that are at least a factor of 2-3 better than current constraints from the CMB \citep{Poulin16}.

As a further step, we ran the same MCMC analysis, this time assuming mock data based on a $\Lambda$CDM universe with massive neutrinos (of $\Sigma m_{\nu}=0.1$ eV) and adding the sum of the neutrino mass as a free model parameter. While this is a more realistic scenario, it does not allow for a direct comparison to previous constraints where neutrinos were assumed to be mass-less. The forecast analysis with massive neutrinos shows similar results compared to the mass-less case with only a $\sim 5$ percent degradation of the DDM parameter constraints. Assuming a scenario where all DM is allowed to decay ($f=1$), we forecast a minimum decay time of $\tau\geq182.23$ Gyr and $\tau\geq275$ Gyr, respectively. The former value corresponds to the pessimistic case where baryonic parameters are marginalised over; the latter value shows the optimistic scenario where baryonic parameters are fixed. Assuming the limit where only a small fraction of DM is allowed to rapidly decay ($\tau<34$ Gyr), we show that WL+CMB data will be able to rule-out decaying-to-total DM fractions above $f=0.21$ and $f=0.13$, respectively, depending on whether baryonic parameters are marginalised over or fixed.

We conclude that upcoming weak-lensing surveys, such as Euclid or LSST at the Rubin Observatory, will allow us to significantly improve upon current constraints for the one-body decaying DM scenario. Compared to current limits from the CMB, an improvement of about a factor of 2-3 can be expected. However, this will require additional knowledge regarding the effects of baryonic feedback on the clustering at cosmological scales.

\acknowledgments
This study is supported by the Swiss National Science Foundation via the grant {\tt PCEFP2\_181157}.

\appendix

\section{CMB priors for the decaying dark matter parameters.}
\label{sec:APoulin}
\begin{figure}[!htb]
\centering 
\includegraphics[width=0.5\textwidth,trim=0.2cm 0.4cm 0.3cm 0.1cm,clip]{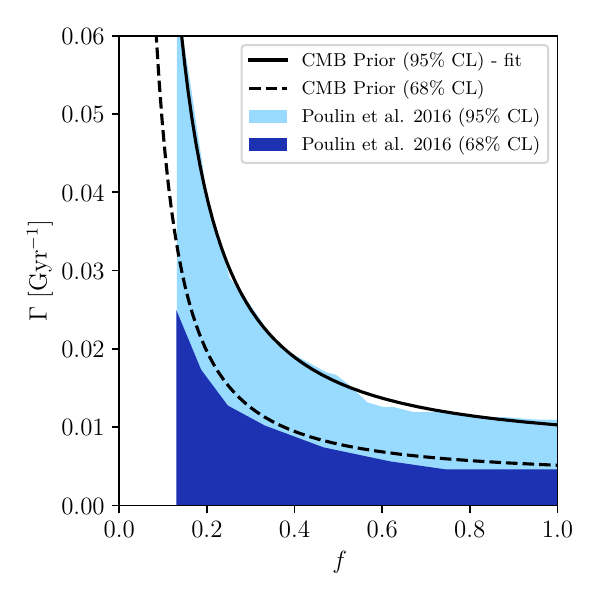}
\caption{Results from Poulin et al. 2016 \citep{Poulin16} (shown in light and dark blue at 95 percent CL and 68 percent CL respectively) and our constructed prior on $\Gamma f^{c_3}$ (68 percent and 95 percent levels shown in dashed and solid lines).}
\label{fig:PoulinComparison}
\end{figure}

In order to assume CMB priors for the DDM parameters in our forecast analysis, we define a prior for the parameter combination $\Gamma f^{c_3}$, where $c_3$ is a constant. We emphasize that in the MCMC analysis, the two parameters $\Gamma$ and $f$ are separate, but we apply a single prior on the combination $\Gamma f^{c_3}$. Starting from the constraints obtained in Ref.~\citep{Poulin16} reported here in Fig.~\ref{fig:PoulinComparison} in blue, we find a fit for the 95 percent CL contour with form $\Gamma = c_1 + c_2/f^{c_3}$ (shown in Fig.~\ref{fig:PoulinComparison} in solid black). Having obtained the fit, we can check the 68 percent CL level by replacing $c_1,\ c_2$ with $c_1/2,\ c_2/2$ (shown in dashed black in Fig.~\ref{fig:PoulinComparison}). We notice that in our construction at the 68 percent CL our fit is more conservative than the results from Ref.\citep{Poulin16}. The 68 percent CL line will therefore be the standard deviation of the Gaussian prior for the parameter combination $\Gamma f^{c_3}$ with mean value 0. Overall, this creates a likelihood distribution which is peaked when $f=0$ or $\Gamma \xrightarrow[]{}0$ and then falls as shown by the 68 percent CL and 95 percent CL contours shown in Fig.~\ref{fig:PoulinComparison}. The values of the fit constants are $(c_1,\ c_2,\ c_3)=(0.00721714,\ 0.00308819,\ 1.45599761)$.

\section{Convergence tests and comparison with the \texorpdfstring{$N$}{N}-body code {\tt CONCEPT}}
\label{sec:AConcept}
\begin{figure}[tb]
\centering 
\subfigure{\includegraphics[width=.32\textwidth,trim=0.3cm 0.4cm 0.4cm 0.4cm,clip]{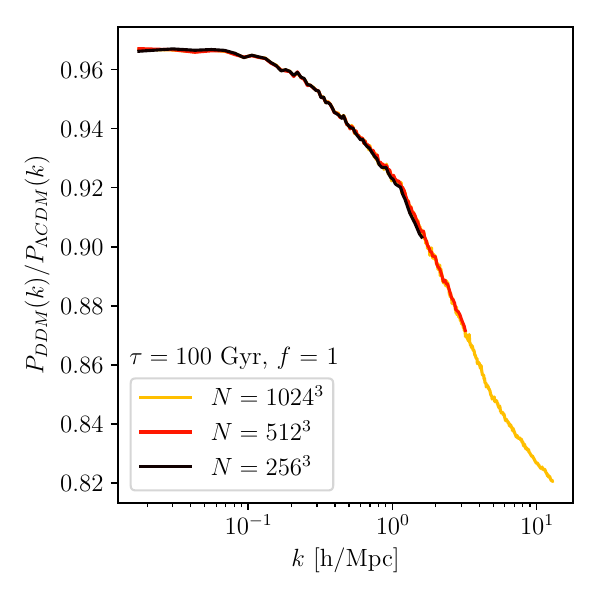}}
\subfigure{\includegraphics[width=.32\textwidth,trim=0.3cm 0.4cm 0.4cm 0.4cm,clip]{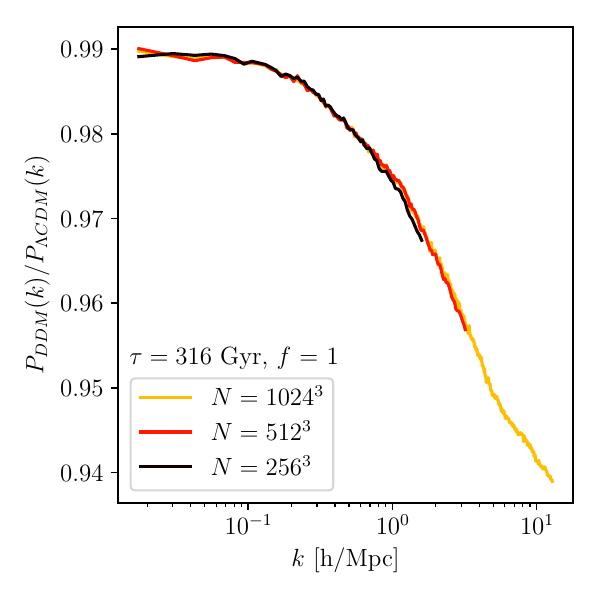}}
\includegraphics[width=0.32\textwidth,trim=0.3cm 0.4cm 0.4cm 0.4cm,clip]{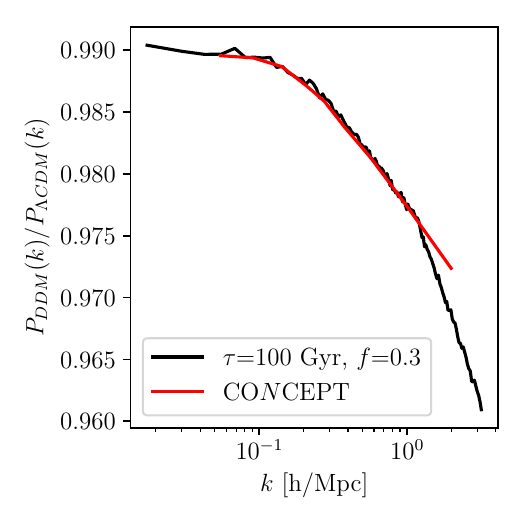}
\caption{Matter power spectrum ratio $P_{DDM}(k)/P_{\Lambda CDM}(k)$ for N-body simulations ran with \texttt{Pkdgrav3} with $N=256^3,\ 512^3,\ 1024^3$ particles for decay lifetimes $\tau = 316 \ \text{Gyr}$, $\tau = 100 \ \text{Gyr}$ and a fixed fraction $f=1$ (left and center panel). In the right panel we show results from \citep{concept} ({\tt CONCEPT}, solid red) and our simulations run in \texttt{Pkdgrav3} (solid black) with $\tau$=100 Gyr and $f$=0.3, 512$^3$ particles and a simulation box size of 500 Mpc/h.}
\label{fig:convergence}
\end{figure}

In this appendix we investigate the convergence of the simulations as well as comparing our results to another $N$-body code, {\tt CONCEPT} \citep{concept}. In order to investigate the convergence properties of our simulations, we show the DDM to CDM power ratio for three simulations with different particle numbers in the left-hand and central panel of Fig.~\ref{fig:convergence}. We thereby limit ourselves to the two cases of $\tau=100$ (left) and $316$ Gyr (right) both with $f=1$. The power ratios from the three different simulations are well converged agreeing at the sub-percent level at least until the Nyquist frequency ($\pi N^{1/3}/L$). Note that this is only true for the ratio plot but not for the absolute power spectrum, where higher resolutions and larger box-sizes are required.

A further test of our $N$-body code and the simulation runs is presented in the right-hand panel of Fig.~\ref{fig:convergence}. There we compare the DDM-to-CDM power spectrum ratio for a model with $\tau=100$ Gyr and $f=0.3$ to the published results from Ref.~\citep{concept} based on the $N$-body code {\tt CONCEPT}. We find sub-percent agreement between the results from the two codes at wave-modes below $k=1$ h/Mpc. The small deviations beyond that scale is most likely due to limited resolution of the simulation with {\tt CONCEPT}.

\section{Forecast posterior distributions.}
\label{sec:AForecast}
\begin{figure}[tb]
\centering 
\includegraphics[width=\textwidth,trim=0.2cm 0.3cm 0.9cm 0.9cm,clip]{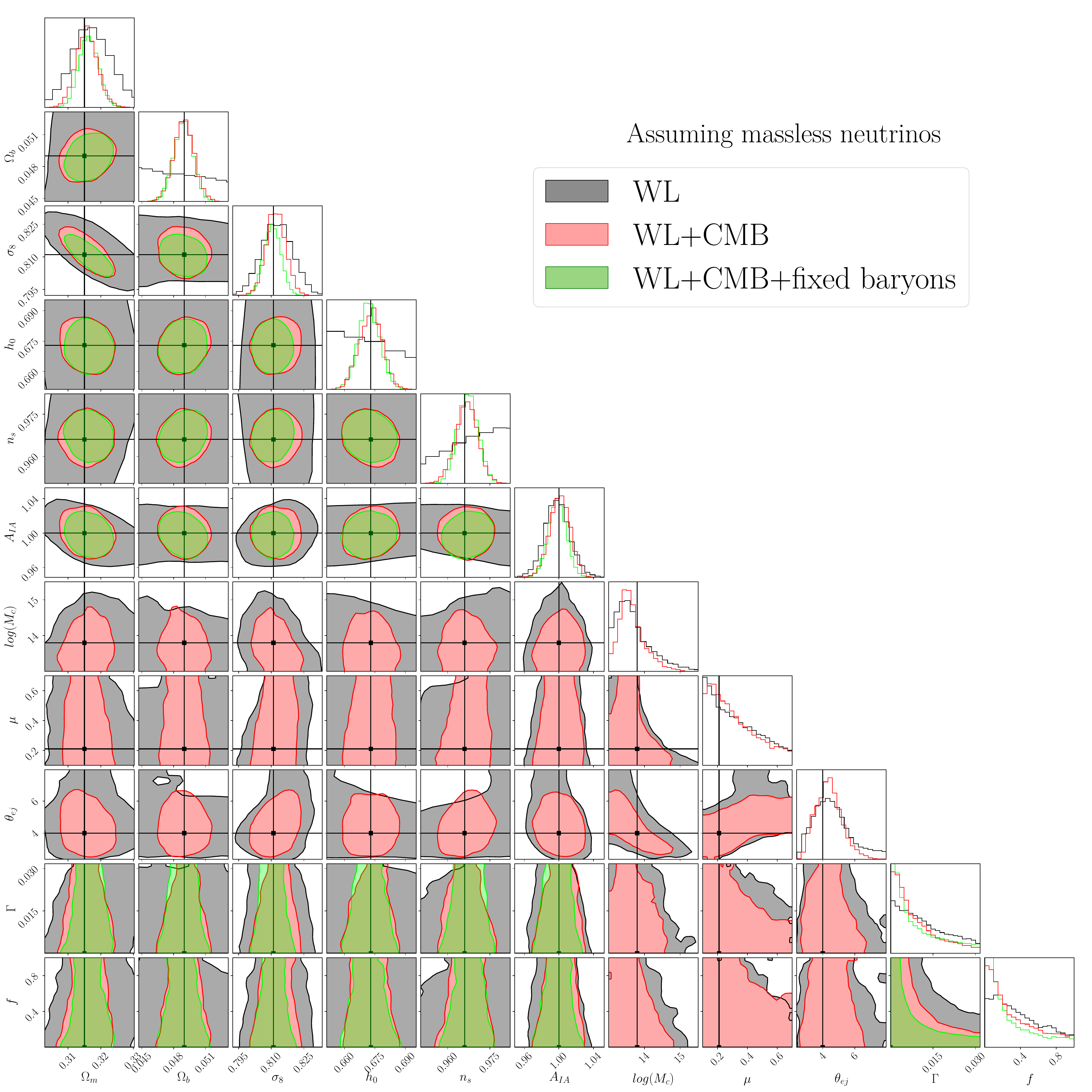}
\caption{Posterior distributions for all the parameters in the MCMC runs when assuming massless neutrinos. In black we show results for the run assuming only WL observables ({WL}), in red we additionally assume CMB priors ({WL+CMB}) from Planck and in green we additionally fix the baryonic parameters ($log(M_c),\ \mu,\ \theta_{ej}$) to their mock value ({WL+CMB+fixed baryons}). All contours are shown at 95 percent CL.}
\label{fig:ForecastWL}
\end{figure}

\begin{figure}[tb]
\centering 
\includegraphics[width=\textwidth,trim=0.1cm 0.3cm 0.9cm 0.9cm,clip]{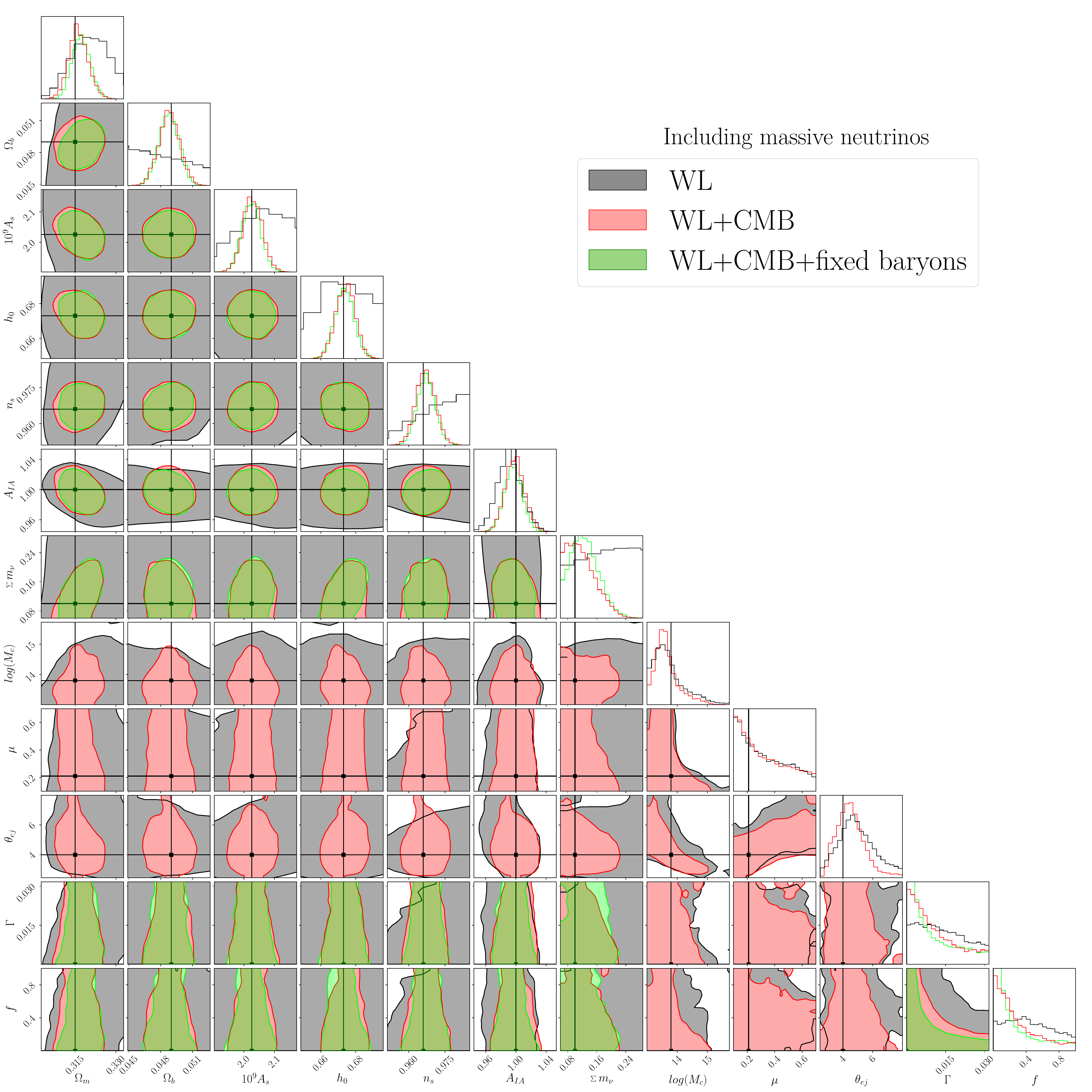}
\caption{Posterior distributions for all the parameters in the MCMC runs when assuming massive neutrinos. In black we show results for the run assuming only WL observables ({WL}), in red we additionally assume CMB priors ({WL+CMB}) from Planck and in green we additionally fix the baryonic parameters ($log(M_c),\ \mu,\ \theta_{ej}$) to their mock value ({WL+CMB+ fixed baryons}). All contours are shown at 95 percent CL.}
\label{fig:ForecastWLn}
\end{figure}

In this section we present the posterior distributions of all the forecast parameters given by our MCMC runs. Fig.~\ref{fig:ForecastWL} and \ref{fig:ForecastWLn} are the corner plots of posterior distributions for the cases without and with massive neutrinos. See Sec.~\ref{sec:forecast} for description of the parameters explored.
The grey, red, and green contours in both figures correspond to the WL only, and the WL+CMB scenarios with free and fixed baryonic parameters (all at 95 percent CL).

We see that the CMB priors improve the constraints substantially for both cases. The baryonic feedback processes can provide similar suppression in the power spectrum as done by the DM decay. Therefore we observe the DDM parameters ($\Gamma$ and $f$) are quite degenerate with the baryonification parameters ($log(M_c),\ \mu,\ \theta_{ej}$). When we fix the baryonification parameters, the constraints on the DDM parameters improves. The improvement is more visible in the $\Gamma$-$f$ plane. See Sec.~\ref{sec:ForecastMassless} and Sec.~\ref{sec:ForecastMassive} in the main text for more discussion.

\bibliographystyle{unsrt}
\bibliography{bibliography}
\end{document}